\documentclass[aps,prl,twocolumn,superscriptaddress,showpacs]{revtex4}
\usepackage{graphicx}

\begin{document}

\def\bax{Ba(Fe$_{1-x}$Co$_{x}$)$_{2}$As$_{2}$}
\def\ba{BaFe$_{2}$As$_{2}$}
\def\tc{$T_{c}$}
\def\rfesi{$R_{2}$Fe$_{3}$Si$_{5}$}
\def\scfesi{Sc$_{2}$Fe$_{3}$Si$_{5}$}
\def\yfesi{Y$_{2}$Fe$_{3}$Si$_{5}$}
\def\tmfesi{Tm$_{2}$Fe$_{3}$Si$_{5}$}
\def\erfesi{Er$_{2}$Fe$_{3}$Si$_{5}$}
\def\etal{{\it et al.}}
\def\spm{$s_{\pm}$}
\def\spp{$s_{++}$}

% Use the \preprint command to place your local institutional report
% number in the upper righthand corner of the title page in preprint mode.
% Multiple \preprint commands are allowed.
% Use the 'preprintnumbers' class option to override journal defaults
% to display numbers if necessary
%\preprint{}

%Title of paper
\title{Suppression of critical temperature in {\bax} with point defects introduced by proton irradiation}

% repeat the \author .. \affiliation  etc. as needed
% \email, \thanks, \homepage, \altaffiliation all apply to the current
% author. Explanatory text should go in the []'s, actual e-mail
% address or url should go in the {}'s for \email and \homepage.
% Please use the appropriate macro foreach each type of information

% \affiliation command applies to all authors since the last
% \affiliation command. The \affiliation command should follow the
% other information
% \affiliation can be followed by \email, \homepage, \thanks as well.
\author{Y.~Nakajima}
\affiliation{Department of Applied Physics, The University of Tokyo, Hongo, Bunkyo-ku, Tokyo 113-8656, Japan}
\affiliation{JST, Transformative Research-Project on Iron Pnictides (TRIP), 7-3-1 Hongo, Bunkyo-ku, Tokyo 113-8656, Japan}
\author{T.~Taen}
\author{Y.~Tsuchiya}
\affiliation{Department of Applied Physics, The University of Tokyo, Hongo, Bunkyo-ku, Tokyo 113-8656, Japan}
\author{T.~Tamegai}
\affiliation{Department of Applied Physics, The University of Tokyo, Hongo, Bunkyo-ku, Tokyo 113-8656, Japan}
\affiliation{JST, Transformative Research-Project on Iron Pnictides (TRIP), 7-3-1 Hongo, Bunkyo-ku, Tokyo 113-8656, Japan}
\author{H.~Kitamura}
\author{T.~Murakami}
\affiliation{Radiation Measurement Research Section, National Institute of Radiological Sciences, 4-9-1, Anagawa, Inage-ku, Chiba 263-8555, Japan}

%\homepage[]{Your web page}
%\thanks{}
%\altaffiliation{}

%Collaboration name if desired (requires use of superscriptaddress
%option in \documentclass). \noaffiliation is required (may also be
%used with the \author command).
%\collaboration can be followed by \email, \homepage, \thanks as well.
%\collaboration{}
%\noaffiliation

\date{\today}

\begin{abstract}
% insert abstract here
We report the effect of 3 MeV proton irradiation on the suppression of the critical temperature $T_{c}$ in Ba(Fe$_{1-x}$Co$_{x}$)$_{2}$As$_{2}$ single crystals at under-, optimal-, and over-doping levels. We find that $T_{c}$ decreases and residual resistivity increases monotonically with increasing dose. We also find no upturn in low-temperature resistivity in contrast with the $\alpha$-particle irradiated NdFeAs(O,F), which suggests that defects induced by the proton irradiation behave as nonmagnetic scattering centers. The critical scattering rate for all samples estimated by three different ways is much higher than that expected in $s_{\pm}$-pairing scenario based on inter-band scattering due to antiferro-magnetic spin fluctuation.
\end{abstract}

% insert suggested PACS numbers in braces on next line
% insert suggested keywords - APS authors don't need to do this
%\keywords{}

%\maketitle must follow title, authors, abstract, \pacs, and \keywords
\maketitle

% body of paper here - Use proper section commands
% References should be done using the \cite, \ref, and \label commands
% Put \label in argument of \section for cross-referencing
%\section{\label{}}

%%\section{Introduction}
Since the discovery of the high-$T_{c}$ iron-based superconductors \cite{kamih08}, extensive studies for the superconducting gap structure have been performed because the gap structure is closely associated with the pairing mechanism. Theoretically, fully-gapped $s$-wave state with opposite signs between different Fermi surfaces ({\spm}-wave) has been proposed \cite{mazin08,kurok08}. The fully-opened gap is suggested by some experiments, such as  penetration depth measurements by microwave conductivity\cite{hashi09,hashi09a}, ARPES \cite{nakay09,teras09a}, and thermal conductivity \cite{luo09}. However, whether the sign-reversal is involved in the multi-gap structure is still controversial. While the inelastic neutron measurements suggests {\it resonance peak} in magnetic excitation spectra $\chi^{''}(\vec{Q},\omega)$, which is observed when the sign of the gap takes opposite values on different parts of the Fermi surface \cite{chris08, inoso10}, it is pointed out that such a peak can be also explained by even {\spp}-symmetry, which has the same sign of gaps on different Fermi surfaces \cite{onari10}. In addition, several studies on the impurity effect indicate that the critical temperature $T_{c}$ is robust against the introduction of non-magnetic impurities, which is strikingly different from the suppression of $T_{c}$ predicted in the {\spm}-wave \cite{lee10,trope10,taran10}. Hence, these results lead us to consider that {\spp}-symmetry should be added to one of the possible candidates for the gap symmetry of iron-based superconductors. Moreover, recent studies suggest that in some iron-based superconductors, such as LaFePO \cite{hicks09}, KFe$_{2}$As$_{2}$ \cite{hashi10b}, and BaFe$_{2}$(As,P)$_{2}$ \cite{hashi10a}, the gap is nodal in a part of the Fermi surface. In Co-doped {\ba}, it is reported that a fully opened gap structure changes to a nodal one when the Co-concentration increases from optimal to over-doped region \cite{reid10}. Further studies on the gap structure of iron-based superconductors have been desired.

To elucidate the superconducting gap structure, a detailed study on the effect of defects is very crucial because the pair-breaking effects due to scattering centers are phase-sensitive. The conventional way to introduce impurities is chemical substitutions of constituent elements. It is well known that isotropic $s$-wave superconductivity is robust against non-magnetic impurities due to Anderson's theorem while superconductivity with a sign change in the gap, such as $d$-wave, is sensitive to non-magnetic impurities. However, chemical substitutions may lead to inhomogeneity in the sample, change of carrier density and the Fermi surface topology, which can mask the intrinsic impurity effect. Another way to introduce scattering centers is to create defects by the swift particle irradiation. Among them, a light element irradiation, such as proton and $\alpha$- particle, is very suitable for the study of artificially introduced scattering centers, since the irradiation can introduce point defects without providing inhomogeneity and changing electronic structure. So far, for single crystals, only one group reports $\alpha$-particle irradiation experiments \cite{taran10}. However, the irradiation produces Kondo-like upturn in the resistivity due to spin flip scattering, which can mask the intrinsic non-magnetic scattering effect on the superconducting gap of iron-arsenide superconductors.

In this letter, we address the issue of the superconducting gap structure in {\bax} ($x$ = 0.045, 0.075, and 0.113) by detailed study of pair-breaking effect introduced by proton irradiation. We find monotonic increase of the resistivity with proton irradiation. The upturn of resistivity at low temperatures is not observed, which indicates that proton irradiation provides non-magnetic scattering centers. The suppression of $T_{c}$ is weaker than the expectation for a superconductor with sign-reversed gaps on/between the Fermi surfaces.

\begin{figure}[t]
\includegraphics[width=8cm]{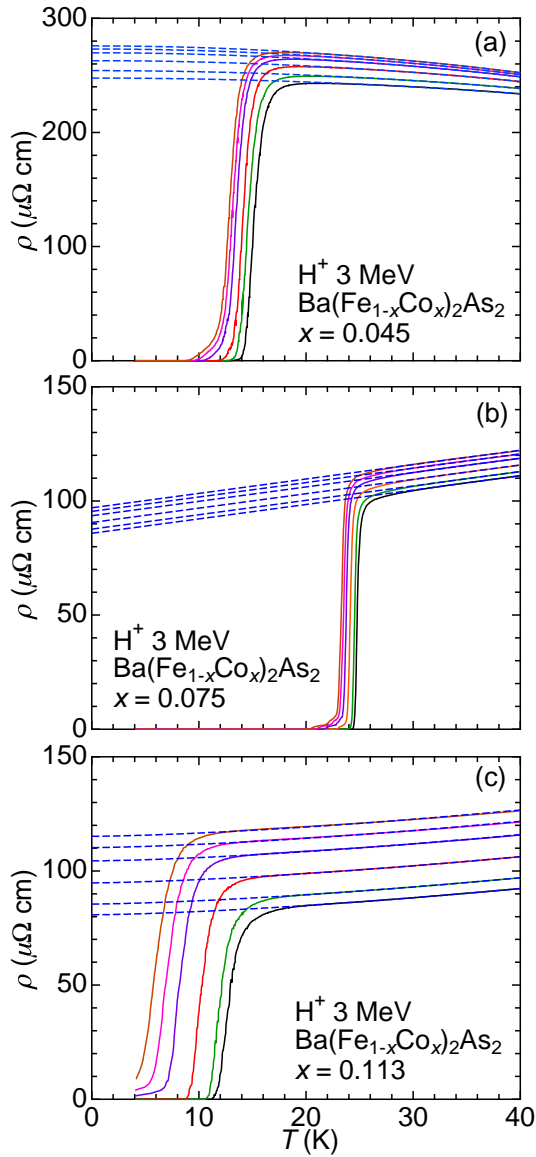}
\caption{(color online) Temperature dependence of the resistivity for {\bax} with (a) $x$ = 0.045, (b) 0.075, and (c) 0.113. The doses are 0, 0.1, 0.5, 0.8, 1.0, and 1.2$\times$10$^{16}$ cm$^{-2}$ from the lowest curve. The dashed blue lines are fit to the data using the equation $\rho=\rho_{0}+AT^{\alpha}$ with $\alpha$ = 2, 1, and 1.5 for $x$ = 0.045, 0.075, and 0.113, respectively. \label{FIG1}}
\end{figure}

\begin{figure}[t]
\includegraphics[width=8cm]{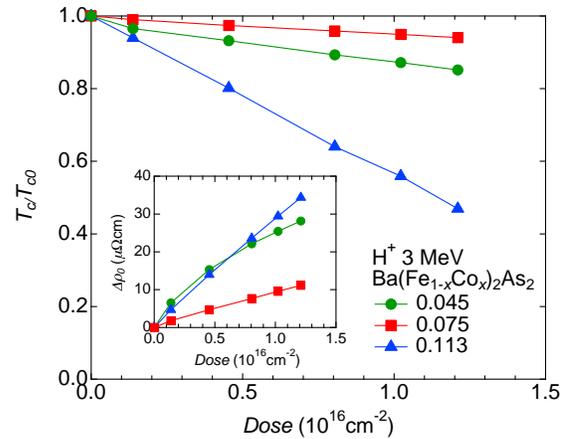}
\caption{(color online) Dose dependence of the normalized critical temperature $T_{c}/T_{c0}$ for {\bax} with (a)$x$ = 0.045, (b) 0.075, and (c) 0.113. The $T_{c0}$ is 15.1, 24.8, and 12.8 K for $x=0.045$, 0.075, and 0.113, respectively. Inset: Increased residual resistivity by irradiation, $\Delta\rho_{0}=\rho_{0}^{irr}-\rho_{0}^{unirr}$, as a function of dose. \label{FIG2}}
\end{figure}

%%%\section{Experimental}

Single-crystalline samples of {\bax} were grown by the FeAs/CoAs self-flux method and their fundamental properties are reported in Ref [\cite{nakaj09a}]. 
%FeAs and CoAs were prepared by placing mixtures of As pieces and Fe/Co powder in a silica tube and reacting them at 1065 $^{\circ}$C for 10 h after heating at 700 $^{\circ}$ C for 6 h. 
A mixture with a ratio of Ba : FeAs/CoAs  = 1 :  5 was  placed in an alumina crucible. The whole assembly was sealed in a large silica tube, and heated up to 1150 $^{\circ}$C and kept there for 10 h followed by slow cooling down to 800 $^{\circ}$C at a rate of 5 $^{\circ}$C/h, which is slightly different from the synthesis reported before \cite{nakaj09}. After cleaving, we can obtain shiny samples. The typical dimensions of the resulting crystals are 4$\times$4$\times$0.1 mm$^{3}$. The average Co concentration in each batch was determined by energy dispersive X-ray spectroscopy measurements. The 3 MeV protons, which are known to create from one to few tens of displacements \cite{cival90}, were irradiated into the samples at HIMAC-NIRS. The irradiation were carried out at 40 K avoiding the thermal annealing effect \cite{xiong88}. A total dose is 1.2 $\times 10^{16}$ cm$^{-2}$. To ensure the uniformity of  damage throughout the sample, we used samples with thicknesses of 15$-$30 $\mu$m, which is smaller than the projected range of $\sim$50 $\mu$m obtained from the simulation using the Stopping and Range of Ions in Matter-2008 \cite{srim}. Resistivity measurements were performed {\it in-situ} after each irradiation by standard four-probe configuration.  Similar results are confirmed in 6 MeV proton irradiation.

%%\section{Results}

Figure 1 shows the resistivity of {\bax} with $x$ = 0.045, 0.075, and 0.113 as a function of temperature. With increasing dose, $T_{c}$ decreases monotonically without significant broadening of the transition width, while the resistivity increases monotonically. It should be noted that in the $\alpha$-particle irradiated NdFeAs(O,F) Kondo-like resistivity upturn at low temperatures is reported \cite{taran10}, which is associated with the spin-flip scattering due to magnetic impurities. In contrast to the behavior in NdFeAs(O,F), no upturn is observed in the resistivity of proton irradiated {\bax}, which strongly suggests that defects produced by the irradiation act as non-magnetic scattering centers.  {\it We emphasize that only the contribution of non-magnetic scattering centers to the pair breaking enable us to investigate the intrinsic non-magnetic impurity effect on the order parameters of iron-arsenide superconductors.}

Figure 2 shows the dose dependence of the normalized critical temperature $T_{c}/T_{c0}$ for {\bax}, where $T_{c0}$ is the transition temperature before the irradiation. $T_{c0}$ obtained from the midpoint of resistive transition are 15.1, 24.8, and 12.8 K for $x$ = 0.045, 0.075, and 0.113, respectively. In all the samples, $T_{c}/T_{c0}$ decreases linearly with increasing the dose in the present dose range. We note that the maximum dose of $\sim 1.2\times 10^{16}$ cm$^{-2}$ in the present study is one forth of that in the $\alpha$-particle irradiated NdFeAs(O,F). Interestingly, in the under- and optimally doped samples, the suppression of $T_{c}$ is very small while in the over-doped sample the suppression is large down to the half of $T_{c0}$. We note that stronger suppressions of $T_{c}$ in overdoped samples are also reported in Zn-doped LaFeAsO$_{0.85}$F$_{0.15}$ \cite{li10} and LaFeAsO$_{0.85}$ \cite{guo10}. The increased residual resistivity by the irradiation $\Delta\rho_{0}$ as a function of dose is plotted in the inset of Fig. 2. $\Delta\rho_{0}$ is the difference of the residual resistivity between irradiated and unirradiated one, namely, $\Delta\rho_{0} =\rho_{0}^{irr}-\rho_{0}^{unirr}$, which corresponds to the density of defects introduced by the proton irradiation. We evaluate the residual resistivity $\rho_{0}$ by fitting the data using $\rho=\rho_{0}+AT^{\alpha}$, where $\alpha$ is an exponent of temperature. We fix $\alpha$ as 2, 1, and 1.5 for $x=0.045$, 0.075, and 0.113, respectively. $\Delta\rho_{0}$ increases almost linearly with dose, which ensures that the proton irradiation introduces defects systematically.

%\begin{figure}[t]
%\includegraphics[width=8cm]{FIG3old.eps}
%\caption{(color online) (a)Normalized critical temperature $T_{c}/T_{c0}$ as a function of normalized scattering rate $g=\hbar\Delta\rho_{0}e/2\pi k_{B}T_{c0}m^{\ast}R_{H}$ for {\bax} with $x$ = 0.045,  0.075, and 0.113. Dashed lines are linear extrapolations. $g_{c}$ is the critical scattering rate expected in {\spm}-scenario. \label{FIG3}}
%\end{figure}

%% to estimate the increase of the elastic scattering rate induced by H^{+} irradiation

\begin{figure}[t]
\includegraphics[width=8cm]{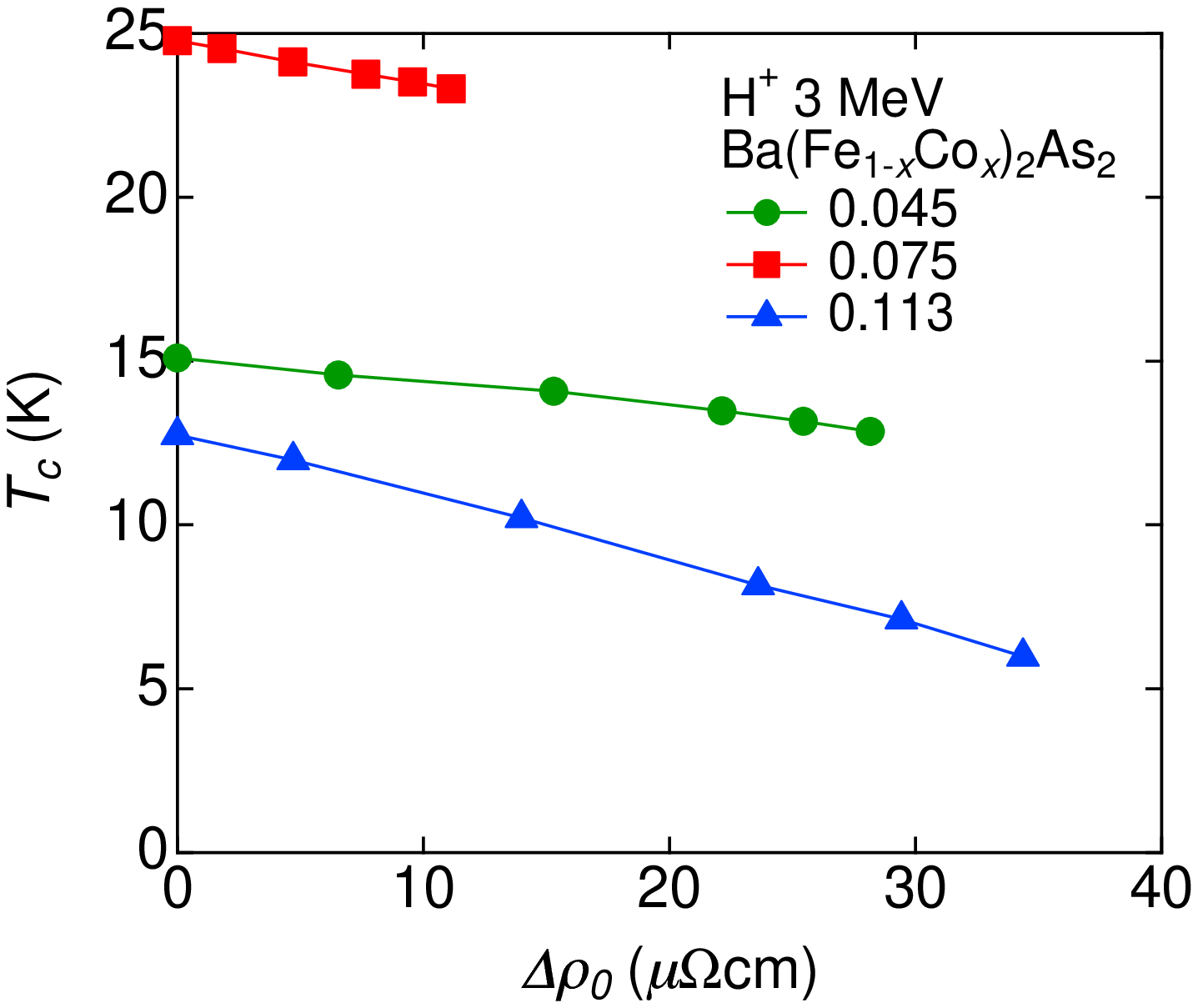}
\caption{(color online) $T_{c}$ as a function of $\Delta\rho_{0}$ for {\bax} with $x$ = 0.045,  0.075, and 0.113. \label{FIG3}}
\end{figure}

\begin{figure}[thb]
\includegraphics[width=7cm]{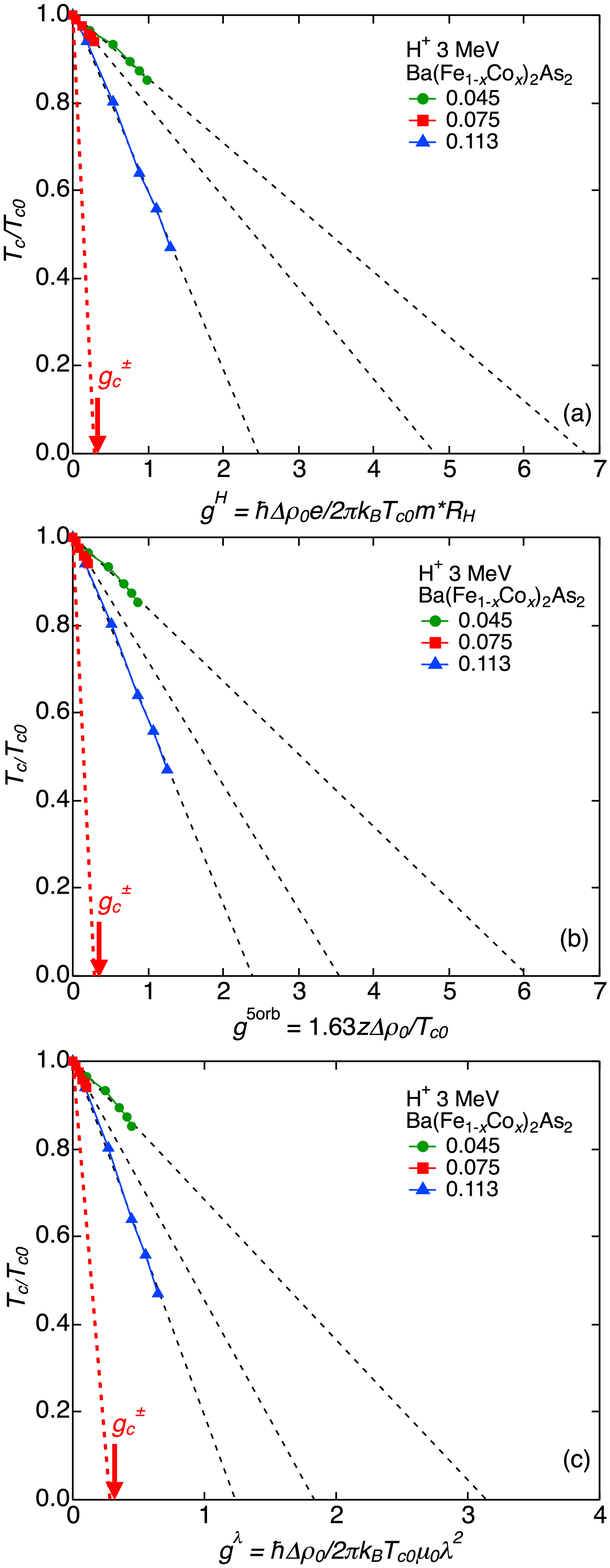}
\caption{(color online) Normalized critical temperature $T_{c}/T_{c0}$ as a function of normalized scattering rate (a) $g^{H}=\hbar\Delta\rho_{0}e/2\pi k_{B}T_{c0}m^{\ast}R_{H}$, (b) $g^{5orb}=1.63z\Delta\rho_{0}/T_{c0}$, and (c) $g^{\lambda}=\hbar\Delta\rho_{0}/2\pi k_{B}T_{c0}\mu_{0}\lambda^{2}$ for {\bax} with $x$ = 0.045,  0.075, and  0.113. Dashed lines are linear extrapolations. $g_{c}^{\pm}$ is the critical scattering rate expected in {\spm}-scenario. \label{FIG4}}
\end{figure}

Figure 3 shows $T_{c}$ as a function of $\Delta\rho_{0}$ for {\bax}. The suppression of $T_{c}$ due to defects introduced by the irradiation is almost linear for samples with all doping levels. The slope $dT_{c}/d(\Delta\rho_{0})$ is $-0.08, -0.13$, and $-0.20$ K/$\mu\Omega$ cm for $x$ = 0.045, 0.075, and 0.113, respectively. These values are slightly larger than the initial slope of the suppression in the $\alpha$-particle irradiated NdFeAs(O,F), $\sim -$ 0.04 K/$\mu\Omega$ cm.

To discuss the pair-breaking effect due to non-magnetic scattering quantitatively, a key parameter is the normalized scattering rate $g=\hbar/2\pi k_{B}T_{c0}\tau$. Here, $\tau$ is the scattering time including both intra- and inter-band scattering contributions. To avoid ambiguity of estimation, we present $g$ obtained from three different ways. In order to obtain the elastic scattering rate introduced by the irradiation, we use the relation $\tau^{-1} = ne^{2}\Delta\rho_{0}/m^{\ast} = e\Delta\rho_{0}/m^{\ast}R_{H}$, where $n$ is the carrier number and  $R_{H}$ is Hall coefficient, $m^{\ast}$ is the effective mass. Figure 4 (a) shows $T_{c}/T_{c0}$ as a function of $g^{H}=\hbar/2\pi k_{B}T_{c0}\tau=\hbar\Delta\rho_{0} e/2\pi k_{B}T_{0}m^{\ast}R_{H}$ for  $x$ = 0.045, 0.075, and 0.113. Assuming that electrons are dominant carriers, we use the $R_{H}$ at 300 K obtained from Ref. \cite{mun09} and $m^{\ast}\sim 3.5m_{e}$ in the electron pocket obtained from ARPES measurements for {\bax} \cite{broue09a}. It should be noted that the estimated $\tau_{0}=m^{\ast}R_{H}/e\rho_{0}$ for unirradated sample with $x$ = 0.075 is $\sim$ 0.02 ps, which is consistent with the scattering time of $\sim$ 0.05 ps just above $T_{c}$ obtained by microwave conductivity measurements for K-doped {\ba} \cite{hashi09} because the residual resistivity for unirradated sample with $x$ = 0.075 is about twice as large as that of K-doped {\ba}.
%It should be noted that the reduction of carrier in  $x$ = 0.045 at low temperatures due to the magnetic transition is not crucial to estimate because even $x$ = 0, the reduction is less than .
According to the {\spm}-scenario with equal gaps of opposite signs on different Fermi surfaces \cite{chubu08}, $T_{c}$ obeys the equation,$-\ln t = \psi(1/2+g/2t)- \psi(1/2)$, where $t=T_{c}/T_{c0}$ and $\psi(x)$ is di-gamma function. This equation indicates that $T_{c}$ vanishes at $g=g_{c}^{\pm}\lesssim 0.3$. To estimate critical values of normalized scattering rate for all samples, we linearly extrapolated the data for simplicity. The obtained values of critical $g_{c}^{H}$ are $\sim$ 6.8, 3.8, and 2.5 for $x=$0.045, 0.075, and 0.113, respectively. Even in $x=$ 0.113, where  $T_{c}$ is most strongly suppressed among them, $g_{c}^{H}$ is much larger than expected $g_{c}^{\pm}$ for $s^{\pm}$-scenario.

We show another way to estimate $g_{c}$ based on the parameter obtained by theoretical calculation. According to the linear response theory based on five orbital model \cite{onari09,konta10}, we can obtain the relation $\Delta \rho_{0}$[$\mu\Omega$cm]$=0.098\tau^{-1}$[K] in {\bax} with interplane distance $c=6.5$ {\AA} and $n=5.8-6.1$.  Figure 4 (b) shows $T_{c}/T_{c0}$ as a function of $g^{5 orb}=z\hbar/2\pi k_{B}T_{c0}\tau=1.63z\Delta \rho_{0}/T_{c0}$, where $z=m/m^{\ast}$ is the renormalization factor. We use $z=1/3.5$ obtained by ARPES \cite{broue09a} assuming $m\sim m_{e}$. Obtained critical values $g_{c}^{5 orb}$ by linear extrapolation are 6.1, 3.5, and 2.4 for $x$ = 0.045, 0.075, and 0.113, respectively, which are again much larger than $g_{c}^{\pm}$ for $s^{\pm}$ scenario. We note that the critical values $g_{c}^{5 orb}$ obtained from the parameter based on theoretical calculation is very similar to $g_{c}^{H}$ obtained from experimental values of $R_{H}$ and $m^{\ast}$.

%\begin{figure}[t]
%\includegraphics[width=8cm]{FIG4old.eps}
%\caption{(color online) Normalized critical temperature $T_{c}/T_{c0}$ as a function of normalized scattering rate $g=1.77z\Delta\rho_{0}/T_{c0}$ for {\bax} with $x$ = 0.045, 0.075, and 0.113. Dashed lines are linear extrapolations. $g_{c}$ is the critical scattering rate expected in {\spm}-scenario.\label{FIG3}}
%\end{figure}

To obtain carrier number and effective mass indirectly, we use the relation $\tau^{-1}=ne^{2}\Delta\rho/m^{\ast} = \Delta\rho/\mu_{0}\lambda^{2}$, where $\lambda$ is the penetration depth, $\lambda=\sqrt{\mu_{0}m^{\ast}/ne^{2}}$. Figure 4 (c) shows  $T_{c}/T_{c0}$ as a function of $g^{\lambda} = \hbar \Delta\rho_{0}/2\pi k_{B}T_{c0}\mu_{0}\lambda^{2}$. Tunnel diode resonator measurements for Al-coated samples provides the absolute values of penetration depths in {\bax} \cite{gordo10a}. Above $x=0.045$, the absolute value of penetration depth is almost independent of Co doping and is close to 200 nm with small scattering. Therefore, we use the value of the penetration depth $\lambda =200$ nm for all samples presented here. Obtained critical values $g_{c}^{\lambda}$ by linear extrapolation are 3.1, 1.8, and 1.2 for $x$ = 0.045, 0.075, and 0.113, respectively. Although these values are roughly half of the previous two estimations, they are more than three times larger than $g_{c}^{\pm}$ expected for $s^{\pm}$-scenario.

%\begin{figure}[t]
%\includegraphics[width=8cm]{FIG5old.eps}
%\caption{(color online) Normalized critical temperature $T_{c}/T_{c0}$ as a function of normalized scattering rate $g=\hbar\Delta\rho_{0}/2\pi k_{B}T_{c0}\mu_{0}\lambda^{2}$ for {\bax} with $x$ = 0.045,  0.075, and  0.113. Dashed lines are linear extrapolations. $g_{c}$ is the critical scattering rate expected in {\spm}-scenario. \label{FIG4}}
%\end{figure}

Critical scattering rates obtained in the three different estimations in the present study are larger than that expected for {\spm} scenario. It should be emphasized that proton irradiation provides only non-magnetic scattering centers without changing electronic structure of {\bax}. Our present results definitely indicates that iron-arsenide superconductor {\bax} is robust against non-magnetic scattering. The weak suppression of $T_{c}$ in {\spm}-wave superconductors could be understood by the details of scattering potential. The pair-breaking effect can be suppressed in very weak or negative scattering potential \cite{onari09}. Strikingly suppressed interband scattering introduced by the irradiation, which is rather unlikely in {\bax} with both electron and hole pockets having $d_{xz}$ and $d_{yz}$ orbital characters, could also explain the weak pair breaking. Further detailed theoretical model to understand the non-trivial weak suppression of $T_{c}$ in {\bax} should be required.

Finally we comment on the possibility of change of the gap structure with doping level in {\bax}. Thermal conductivity measurements suggest the existence of nodes in over-doped sample \cite{reid10}. The stronger suppression of $T_{c}$ in over-doped sample than under- and optimally doped ones may suggest the different gap structure, for instance,  nodal {\spm}-wave symmetry\cite{kurok09a}, where the order parameter has $d$-wave-like nodes on the electron Fermi surface while others are fully open, or {\spm}-wave with {\it accidental} horizontal nodes \cite{reid10}.

In summary, we present the suppression of $T_{c}$ by  defects introduced by 3 MeV proton irradiation in {\bax} single crystals with different doping levels. We find that $T_{c}$ decreases and residual resistivity increases monotonically. No Kondo-like upturn in the low temperature resistivity is observed which suggests that defects created by the irradiation act as non-magnetic scattering centers. The critical scattering rates obtained from the three different estimations are much larger than that expected in {\spm}-scenario, which may contradict the theoretical expectation based on inter-pocket scattering due to antiferromagnetic spin fluctuations.

We thank Y. Matsuda, T. Shibauchi, H. Kontani, and R. Arita for useful discussions. This work is partly supported by a Grant-in-Aid for Scientific Reserch from the MEXT Japan.


\begin{thebibliography}{32}
\expandafter\ifx\csname natexlab\endcsname\relax\def\natexlab#1{#1}\fi
\expandafter\ifx\csname bibnamefont\endcsname\relax
  \def\bibnamefont#1{#1}\fi
\expandafter\ifx\csname bibfnamefont\endcsname\relax
  \def\bibfnamefont#1{#1}\fi
\expandafter\ifx\csname citenamefont\endcsname\relax
  \def\citenamefont#1{#1}\fi
\expandafter\ifx\csname url\endcsname\relax
  \def\url#1{\texttt{#1}}\fi
\expandafter\ifx\csname urlprefix\endcsname\relax\def\urlprefix{URL }\fi
\providecommand{\bibinfo}[2]{#2}
\providecommand{\eprint}[2][]{\url{#2}}

\bibitem[{\citenamefont{Kamihara {\it et~al.}}(2008)\citenamefont{Kamihara, Watanabe,
  Hirano, and Hosono}}]{kamih08}
\bibinfo{author}{\bibfnamefont{Y.}~\bibnamefont{Kamihara}} \bibnamefont{{\it et~al.}},
  \bibinfo{journal}{J. Am. Chem. Soc.} \textbf{\bibinfo{volume}{130}},
  \bibinfo{pages}{3296} (\bibinfo{year}{2008}).

\bibitem[{\citenamefont{Mazin {\it et~al.}}(2008)\citenamefont{Mazin, Singh,
  Johannes, and Du}}]{mazin08}
\bibinfo{author}{\bibfnamefont{I.~I.} \bibnamefont{Mazin}}, 
\bibinfo{author}{\bibfnamefont{D.~J.} \bibnamefont{Singh}},
  \bibinfo{author}{\bibfnamefont{M.~D.} \bibnamefont{Johannes}},
  \bibnamefont{and} \bibinfo{author}{\bibfnamefont{M.~H.} \bibnamefont{Du}},
  \bibinfo{journal}{Phys. Rev. Lett.} \textbf{\bibinfo{volume}{101}},
  \bibinfo{pages}{057003} (\bibinfo{year}{2008}).

\bibitem[{\citenamefont{Kuroki {\it et~al.}}(2008)\citenamefont{Kuroki, Onari, Arita,
  Usui, Tanaka, Kontani, and Aoki}}]{kurok08}
\bibinfo{author}{\bibfnamefont{K.}~\bibnamefont{Kuroki}} \bibnamefont{{\it et~al.}},
  \bibinfo{journal}{Phys. Rev. Lett.} \textbf{\bibinfo{volume}{101}},
  \bibinfo{pages}{087004} (\bibinfo{year}{2008}).

\bibitem[{\citenamefont{Hashimoto
  {\it et~al.}}(2009{\natexlab{a}})\citenamefont{Hashimoto, 
  {\it et~al.}}}]{hashi09}
\bibinfo{author}{\bibfnamefont{K.}~\bibnamefont{Hashimoto}} \bibnamefont{{\it et~al.}}, \bibinfo{journal}{Phys. Rev. Lett.}
  \textbf{\bibinfo{volume}{102}}, \bibinfo{pages}{207001}
  (\bibinfo{year}{2009}{\natexlab{a}}).

\bibitem[{\citenamefont{Hashimoto
  {\it et~al.}}(2009{\natexlab{b}})\citenamefont{Hashimoto, Shibauchi, Kato, Ikada,
  Okazaki, Shishido, Ishikado, Kito, Iyo, Eisaki {\it et~al.}}}]{hashi09a}
\bibinfo{author}{\bibfnamefont{K.}~\bibnamefont{Hashimoto}} \bibnamefont{{\it et~al.}}, \bibinfo{journal}{Phys. Rev. Lett.}
  \textbf{\bibinfo{volume}{102}}, \bibinfo{pages}{017002}
  (\bibinfo{year}{2009}{\natexlab{b}}).

\bibitem[{\citenamefont{Nakayama {\it et~al.}}(2009)\citenamefont{Nakayama, Sato,
  Richard, Xu, Sekiba, Souma, Chen, Luo, Wang, Ding {\it et~al.}}}]{nakay09}
\bibinfo{author}{\bibfnamefont{K.}~\bibnamefont{Nakayama}}
\bibnamefont{{\it et~al.}},
  \bibinfo{journal}{Europhys. Lett.} \textbf{\bibinfo{volume}{85}},
  \bibinfo{pages}{67002} (\bibinfo{year}{2009}).

\bibitem[{\citenamefont{Terashima {\it et~al.}}(2009)\citenamefont{Terashima, Sekiba,
  Bowen, Nakayama, Kawahara, Sato, Richard, Xu, Li, Cao {\it et~al.}}}]{teras09a}
\bibinfo{author}{\bibfnamefont{K.}~\bibnamefont{Terashima}} \bibnamefont{{\it et~al.}}, \bibinfo{journal}{Proc. Natl. Acad. Sci.}
  \textbf{\bibinfo{volume}{106}}, \bibinfo{pages}{7330} (\bibinfo{year}{2009}).

\bibitem[{\citenamefont{Luo {\it et~al.}}(2009)\citenamefont{Luo, Tanatar, Reid,
  Shakeripour, Doiron-Leyraud, Ni, Bud'ko, Canfield, Luo, Wang {\it et~al.}}}]{luo09}
\bibinfo{author}{\bibfnamefont{X.~G.} \bibnamefont{Luo}}
\bibnamefont{{\it et~al.}},
  \bibinfo{journal}{Phys. Rev. B} \textbf{\bibinfo{volume}{80}},
  \bibinfo{pages}{140503} (\bibinfo{year}{2009}).

\bibitem[{\citenamefont{Christianson {\it et~al.}}(2008)\citenamefont{Christianson,
  Goremychkin, Osborn, Rosenkranz, Lumsden, Malliakas, Todorov, Claus, Chung,
  Kanatzidis {\it et~al.}}}]{chris08}
\bibinfo{author}{\bibfnamefont{A.~D.} \bibnamefont{Christianson}} \bibnamefont{{\it et~al.}}, \bibinfo{journal}{Nature}
  \textbf{\bibinfo{volume}{456}}, \bibinfo{pages}{930} (\bibinfo{year}{2008}).

\bibitem[{\citenamefont{Inosov {\it et~al.}}(2010)\citenamefont{Inosov, Park,
  Bourges, Sun, Sidis, Schneidewind, Hradil, Haug, Lin, Keimer
  {\it et~al.}}}]{inoso10}
\bibinfo{author}{\bibfnamefont{D.~S.} \bibnamefont{Inosov}} \bibnamefont{{\it et~al.}}, \bibinfo{journal}{Nat. Phys.}
  \textbf{\bibinfo{volume}{6}}, \bibinfo{pages}{178} (\bibinfo{year}{2010}).

\bibitem[{\citenamefont{Onari {\it et~al.}}(2010)\citenamefont{Onari, Kontani, and
  Sato}}]{onari10}
\bibinfo{author}{\bibfnamefont{S.}~\bibnamefont{Onari}},
  \bibinfo{author}{\bibfnamefont{H.}~\bibnamefont{Kontani}}, \bibnamefont{and}
  \bibinfo{author}{\bibfnamefont{M.}~\bibnamefont{Sato}},
  \bibinfo{journal}{Phys. Rev. B} \textbf{\bibinfo{volume}{81}},
  \bibinfo{pages}{060504} (\bibinfo{year}{2010}).

\bibitem[{\citenamefont{Lee {\it et~al.}}(2010)\citenamefont{Lee, Satomi, Kobayashi,
  and Sato}}]{lee10}
\bibinfo{author}{\bibfnamefont{S.~C.} \bibnamefont{Lee}} \bibnamefont{{\it et~al.}},
  \bibinfo{journal}{J. Phys. Soc. Jpn.} \textbf{\bibinfo{volume}{79}},
  \bibinfo{pages}{023702} (\bibinfo{year}{2010}).

\bibitem[{\citenamefont{Tropeano {\it et~al.}}(2010)\citenamefont{Tropeano, Cimberle,
  Ferdeghini, Lamura, Martinelli, Palenzona, Pallecchi, Sala, Sheikin,
  Bernardini {\it et~al.}}}]{trope10}
\bibinfo{author}{\bibfnamefont{M.}~\bibnamefont{Tropeano}} \bibnamefont{{\it et~al.}}, \bibinfo{journal}{Phys. Rev. B}
  \textbf{\bibinfo{volume}{81}}, \bibinfo{pages}{184504}
  (\bibinfo{year}{2010}).

\bibitem[{\citenamefont{Tarantini {\it et~al.}}(2010)\citenamefont{Tarantini, Putti,
  Gurevich, Shen, Singh, Rowell, Newman, Larbalestier, Cheng, Jia
  {\it et~al.}}}]{taran10}
\bibinfo{author}{\bibfnamefont{C.}~\bibnamefont{Tarantini}} \bibnamefont{{\it et~al.}},
  \bibinfo{journal}{Phys. Rev. Lett.} \textbf{\bibinfo{volume}{104}},
  \bibinfo{pages}{087002} (\bibinfo{year}{2010}).

\bibitem[{\citenamefont{Hicks {\it et~al.}}(2009)\citenamefont{Hicks, Lippman, Huber,
  Analytis, Chu, Erickson, Fisher, and Moler}}]{hicks09}
\bibinfo{author}{\bibfnamefont{C.~W.} \bibnamefont{Hicks}} \bibnamefont{{\it et~al.}},
  \bibinfo{journal}{Phys. Rev. Lett.} \textbf{\bibinfo{volume}{103}},
  \bibinfo{pages}{127003} (\bibinfo{year}{2009}).

\bibitem[{\citenamefont{Hashimoto
  {\it et~al.}}(2010{\natexlab{a}})\citenamefont{Hashimoto, Serafin, Tonegawa,
  Katsumata, Okazaki, Saito, Fukazawa, Kohori, Kihou, Lee {\it et~al.}}}]{hashi10b}
\bibinfo{author}{\bibfnamefont{K.}~\bibnamefont{Hashimoto}} \bibnamefont{{\it et~al.}}, \bibinfo{journal}{Phys. Rev. B}
  \textbf{\bibinfo{volume}{82}}, \bibinfo{pages}{014526}
  (\bibinfo{year}{2010}{\natexlab{a}}).

\bibitem[{\citenamefont{Hashimoto
  {\it et~al.}}(2010{\natexlab{b}})\citenamefont{Hashimoto, Yamashita, Kasahara,
  Senshu, Nakata, Tonegawa, Ikada, Serafin, Carrington, Terashima
  {\it et~al.}}}]{hashi10a}
\bibinfo{author}{\bibfnamefont{K.}~\bibnamefont{Hashimoto}} \bibnamefont{{\it et~al.}}, \bibinfo{journal}{Phys. Rev. B}
  \textbf{\bibinfo{volume}{81}}, \bibinfo{pages}{220501}
  (\bibinfo{year}{2010}{\natexlab{b}}).

\bibitem[{\citenamefont{Reid {\it et~al.}}(2010)\citenamefont{Reid, Tanatar, Luo,
  Shakeripour, Doiron-Leyraud, Ni, Bud'ko, Canfield, Prozorov, and
  Taillefer}}]{reid10}
\bibinfo{author}{\bibfnamefont{J.-P.} \bibnamefont{Reid}} \bibnamefont{{\it et~al.}},
  \bibinfo{journal}{Phys. Rev. B} \textbf{\bibinfo{volume}{82}},
  \bibinfo{pages}{064501} (\bibinfo{year}{2010}).
  
\bibitem[{\citenamefont{Nakajima
  et~al.}(2009{\natexlab{a}})\citenamefont{Nakajima, Taen, and
  Tamegai}}]{nakaj09a}
\bibinfo{author}{\bibfnamefont{Y.}~\bibnamefont{Nakajima}},
  \bibinfo{author}{\bibfnamefont{T.}~\bibnamefont{Taen}}, \bibnamefont{and}
  \bibinfo{author}{\bibfnamefont{T.}~\bibnamefont{Tamegai}},  to be published in 
  \bibinfo{journal}{Physica C}.

\bibitem[{\citenamefont{Nakajima {\it et~al.}}(2009)\citenamefont{Nakajima, Taen, and
  Tamegai}}]{nakaj09}
\bibinfo{author}{\bibfnamefont{Y.}~\bibnamefont{Nakajima}}
 \bibnamefont{{\it et~al.}},
  \bibinfo{journal}{J. Phys. Soc. Jpn.} \textbf{\bibinfo{volume}{78}},
  \bibinfo{pages}{023702} (\bibinfo{year}{2009}).
  
  \bibitem[{\citenamefont{Civale et~al.}(1990)\citenamefont{Civale, Marwick,
  McElfresh, Worthington, Malozemoff, Holtzberg, Thompson, and Kirk}}]{cival90}
\bibinfo{author}{\bibfnamefont{L.}~\bibnamefont{Civale}} \bibnamefont{{\it et~al.}},
  \bibinfo{journal}{Phys. Rev. Lett.} \textbf{\bibinfo{volume}{65}},
  \bibinfo{pages}{1164} (\bibinfo{year}{1990}).

\bibitem[{\citenamefont{Xiong {\it et~al.}}(1988)\citenamefont{Xiong, Li, Linker, and
  Meyer}}]{xiong88}
\bibinfo{author}{\bibfnamefont{G.~C.} \bibnamefont{Xiong}},
 \bibinfo{author}{\bibfnamefont{H.~C.} \bibnamefont{Li}},
  \bibinfo{author}{\bibfnamefont{G.}~\bibnamefont{Linker}}, \bibnamefont{and}
  \bibinfo{author}{\bibfnamefont{O.}~\bibnamefont{Meyer}},
  \bibinfo{journal}{Phys. Rev. B} \textbf{\bibinfo{volume}{38}},
  \bibinfo{pages}{240} (\bibinfo{year}{1988}).

\bibitem[{\citenamefont{Ziegler {\it et~al.}}(1985)\citenamefont{Ziegler, Biersack,
  and Littmark}}]{srim}
\bibinfo{author}{\bibfnamefont{J.~F.} \bibnamefont{Ziegler}},
  \bibinfo{author}{\bibfnamefont{J.~P.} \bibnamefont{Biersack}},
  \bibnamefont{and} \bibinfo{author}{\bibfnamefont{U.}~\bibnamefont{Littmark}},
  \emph{\bibinfo{title}{The Stopping and Range of Ions in Solids}}
  (\bibinfo{publisher}{Pergamon Press}, \bibinfo{address}{New York},
  \bibinfo{year}{1985}).

\bibitem[{\citenamefont{Li {\it et~al.}}(2010)\citenamefont{Li, Tong, Tao, Feng, Cao,
  Chen, chun Zhang, and an~Xu}}]{li10}
\bibinfo{author}{\bibfnamefont{Y.}~\bibnamefont{Li}} \bibnamefont{{\it et~al.}},
  \bibinfo{journal}{New J. Phys.} \textbf{\bibinfo{volume}{12}},
  \bibinfo{pages}{083008} (\bibinfo{year}{2010}).

\bibitem[{\citenamefont{Guo {\it et~al.}}(2010)\citenamefont{Guo, Shi, Yu, Belik,
  Matsushita, Tanaka, Katsuya, Kobayashi, Nowik, Felner {\it et~al.}}}]{guo10}
\bibinfo{author}{\bibfnamefont{Y.~F.} \bibnamefont{Guo}},
   \bibnamefont{{\it et~al.}}, \bibinfo{journal}{Phys. Rev. B}
  \textbf{\bibinfo{volume}{82}}, \bibinfo{pages}{054506}
  (\bibinfo{year}{2010}).

\bibitem[{\citenamefont{Mun {\it et~al.}}(2009)\citenamefont{Mun, Bud'ko, Ni, Thaler,
  and Canfield}}]{mun09}
\bibinfo{author}{\bibfnamefont{E.~D.} \bibnamefont{Mun}},
 \bibinfo{author}{\bibfnamefont{S.~L.} \bibnamefont{Bud'ko}},
  \bibinfo{author}{\bibfnamefont{N.}~\bibnamefont{Ni}},
  \bibinfo{author}{\bibfnamefont{A.~N.} \bibnamefont{Thaler}},
  \bibnamefont{and} \bibinfo{author}{\bibfnamefont{P.~C.}\bibnamefont{Canfield}},
 \bibinfo{journal}{Phys. Rev. B}
  \textbf{\bibinfo{volume}{80}}, \bibinfo{eid}{054517}
 (\bibinfo{year}{2009}).

\bibitem[{\citenamefont{Brouet {\it et~al.}}(2009)\citenamefont{Brouet, Marsi,
  Mansart, Nicolaou, Taleb-Ibrahimi, Le~F\`evre, Bertran, Rullier-Albenque,
  Forget, and Colson}}]{broue09a}
\bibinfo{author}{\bibfnamefont{V.}~\bibnamefont{Brouet}} \bibnamefont{{\it et~al.}},
  \bibinfo{journal}{Phys. Rev. B} \textbf{\bibinfo{volume}{80}},
  \bibinfo{pages}{165115} (\bibinfo{year}{2009}).

\bibitem[{\citenamefont{Chubukov {\it et~al.}}(2008)\citenamefont{Chubukov, Efremov,
  and Eremin}}]{chubu08}
\bibinfo{author}{\bibfnamefont{A.~V.} \bibnamefont{Chubukov}},
  \bibinfo{author}{\bibfnamefont{D.~V.} \bibnamefont{Efremov}},
  \bibnamefont{and} \bibinfo{author}{\bibfnamefont{I.}~\bibnamefont{Eremin}},
  \bibinfo{journal}{Phys. Rev. B} \textbf{\bibinfo{volume}{78}},
  \bibinfo{pages}{134512} (\bibinfo{year}{2008}).

\bibitem[{\citenamefont{Onari and Kontani}(2009)}]{onari09}
\bibinfo{author}{\bibfnamefont{S.}~\bibnamefont{Onari}} \bibnamefont{and}
  \bibinfo{author}{\bibfnamefont{H.}~\bibnamefont{Kontani}},
  \bibinfo{journal}{Phys. Rev. Lett.} \textbf{\bibinfo{volume}{103}},
  \bibinfo{pages}{177001} (\bibinfo{year}{2009}).

\bibitem[{\citenamefont{Kontani and Sato}(unpublished.)}]{konta10}
\bibinfo{author}{\bibfnamefont{H.}~\bibnamefont{Kontani}} \bibnamefont{and}
  \bibinfo{author}{\bibfnamefont{M.}~\bibnamefont{Sato}},
  \bibinfo{journal}{arXiv:1005.0942}  (\bibinfo{year}{unpublished}).

\bibitem[{\citenamefont{Gordon {\it et~al.}}(2010)\citenamefont{Gordon, Kim,
  Salovich, Giannetta, Fernandes, Kogan, Prozorov, Bud'ko, Canfield, Tanatar
  {\it et~al.}}}]{gordo10a}
\bibinfo{author}{\bibfnamefont{R.~T.} \bibnamefont{Gordon}} \bibnamefont{{\it et~al.}}, \bibinfo{journal}{Phys. Rev. B}
  \textbf{\bibinfo{volume}{82}}, \bibinfo{pages}{054507}
  (\bibinfo{year}{2010}).

\bibitem[{\citenamefont{Kuroki {\it et~al.}}(2009)\citenamefont{Kuroki, Usui, Onari,
  Arita, and Aoki}}]{kurok09a}
\bibinfo{author}{\bibfnamefont{K.}~\bibnamefont{Kuroki}},
  \bibinfo{author}{\bibfnamefont{H.}~\bibnamefont{Usui}},
  \bibinfo{author}{\bibfnamefont{S.}~\bibnamefont{Onari}},
  \bibinfo{author}{\bibfnamefont{R.}~\bibnamefont{Arita}}, \bibnamefont{and}
  \bibinfo{author}{\bibfnamefont{H.}~\bibnamefont{Aoki}},
  \bibinfo{journal}{Phys. Rev. B} \textbf{\bibinfo{volume}{79}},
  \bibinfo{pages}{224511} (\bibinfo{year}{2009}).

\end{thebibliography}
\end{document}